\documentstyle[12pt]{article}
\def\fnote#1#2{\begingroup\def\thefootnote{#1}\footnote{#2}
\endgroup}

\begin{document}
\vspace{36pt}

\begin{center}
{\large{\bf Cosmic Rays from Decaying Vortons}}

\vspace{36pt}
Luis Masperi\fnote{*}{E-mail: masperi@cab.cnea.edu.ar}
and Guillermo Silva\fnote{+}{E-mail: silvag@cab.cnea.edu.ar}\\
{\em Centro At\'omico Bariloche and Instituto Balseiro \\
(Comisi\'on Nacional de Energ\'{\i}a At\'omica \\
and Universidad Nacional de Cuyo)\\
 8400 S.C. de Bariloche, Argentina\\}
\end{center}

\vspace{30pt}

\begin{center}

{\bf Abstract}
\end{center}

\noindent
The flux of high energy cosmic rays coming from the decay of vortons is
estimated. If the abundance of loops corresponding to a superconductivity
scale coincident with that of the string formation is corrected to be
compatible with the critical density of universe, it is found that the
emission of one carrier per vorton may produce a flux of one cosmic ray
event per $km^2$ of detector and per year.

\vfill

\baselineskip=20pt
\pagebreak
\setcounter{page}{1}

\begin{center}
{\em I. Introduction}
\end{center}

It is difficult to explain the source of the observed ultra high energy
cosmic rays (UHECR) exceeding a few $10^{19}~eV$ [1].

Standard acceleration mechanisms hardly justify energies higher than
$10^{15}~eV$ [2]. It is possible that UHECR have an extragalactic origin in
AGN, which cannot be however at a distance larger than $100~Mpc$ to avoid
their degrading through pion photoproduction due to scattering with CBR
[3].

For the case that these relatively nearby sources cannot be identified,
an alternative explanation may be a top-down production of UHECR from
Grand Unified Theories (GUT) particles emitted by topological defects like
cosmic strings [4].

However, normal cosmic strings formed at the GUT scale suffer a
dynamics which reduces their density at present in such a degree that the
possible flux of cosmic rays that they may produce is several orders of
magnitude lower than the expected one [5].

Cosmic strings may become more stable if they have a superconducting
core [6]. When they are very long, their motion through the intergalactic
magnetic field induces strong currents which favour the emission of a
high mass carrier whose decay might produce the UHECR. But the extreme
intense magnetic field surrounding the superconducting string would
degrade the particle energy through syncrotron radiation.

A more plausible scenario is that of vortons, small superconducting
closed strings stabilized by their angular momentum [7]. Their present
density is determined classically by the scale at which they acquire the
superconducting property [8] that is constrained by the primordial
nucleosynthesis and the critical density of the universe [9].

In the present work we have estimated the flux of cosmic rays from
quantum decay of vortons. The evaluation is based on the tunneling of a
chiral vorton, with equal topological and charge numbers, to a configuration
with one less unit. It is seen that vortons which acquire superconductivity
at a scale much lower than that of GUT have a negligible probability of
tunneling decay. But if instead superconductivity appeared at the string
formation, quantum decay or other mechanisms may reduce vorton abundance to
become consistent with the universe critical density and a flux of cosmic
rays compatible with the expected one might be obtained.

In Section II we derive the expression of the UHECR
flux from the density of vortons in terms of its lifetime which is
estimated by a semiclassical method in Section III. Conclusions are given
in Section IV.

\vspace{12pt}
\begin{center}
{\em II. The Flux of Cosmic Rays}
\end{center}

\nopagebreak
As it was done in Ref.[5], using conformal time and space
$ds^2=a^2(\tau)(d\tau^2-dx^2)$, the number of events in a spherical
shell during a conformal time interval is

\begin{equation}
<n(\tau)>4\pi a^2 x^2~adx~ad\tau~~~.
\end{equation}

The fraction that will be observed by a detector of area
$A$ is $A/4\pi a^2_0 x^2$, where $a_0$ is the present scale.

If for each event there are $N_c$ produced particles the
present detected flux will be

\begin{equation}
{1 \over A}<{d{\cal N} \over dt_0}>=\int^{t_0}_{t_{eq}} N_c <n(\tau)>
{a^3 \over a_0^3} dt~~~,
\end{equation}

\noindent
where the integration is extended back to the equivalence time between
radiation and matter because beyond it the cosmic rays would have been too
heavily redshifted, since $z_{eq}~\sim~a~few~10^4$, the eventually emitted
GUT particle has a mass $M_X \sim 10^{24}~eV$ and we are interested in UHECR
of energy greater than a few  $10^{19}~eV$.

The number of events per unit volume and time would be related to the
vorton density and lifetime for the decay mode of one carrier by
$<~n(\tau)~>~=~n_v~/~\tau_v$.

If one assumes that the scale for the appearance of superconducting
properties coincides with that of the string formation $M_X \sim
M_{GUT}=\eta$, the vorton density turns out to be classically [9]

\begin{equation}
n_v \sim ({M_X \over m_{pl}})^{3/2}~ T^3~~~,
\end{equation}

\noindent
which is much larger than the critical density of the universe and would
give an enormous cosmic ray flux, unless vortons are extremely stable under
quantum decay. On the other hand if the scale for superconductivity
$m_\sigma$ is lower than $M_X$, the vorton density is reduced to

\begin{equation}
n_v\sim ({m_\sigma \over M_X})^{9/2}~({M_X \over m_{pl}})^{3/2}~ T^3~~~,
\end{equation}

\noindent
when the condensation is produced in the string friction regime. For
$m_\sigma~<~M_X^2/m_{pl}$ the string radiation regime applies but the
consequences are similar. If $m_\sigma~\sim~10^9~GeV$ this vorton abundance
is consistent with the universe critical density.

Taking the matter dominance scaling

\begin{equation}
a/a_0=(t/t_0)^{2/3}~~~,~~~T \sim 10~eV~(t_{eq}/t)^{2/3}~~~,
\end{equation}

\noindent
and for this order of vorton density, the present cosmic ray flux would be

\begin{equation}
{1 \over A}<{d{\cal N} \over {dt_0}}>= {N_c \over \tau_v[yr]}~10^{9}~{1
\over {km^2~yr}}~~~.
\end{equation}

Depending on the vorton details which determine its lifetime for the
relevant decay mode, if it is of the order of the universe age and expecting
$N_c \sim 10$ as the number of UHECR per emitted carrier, Eq.(5) might be
consistent with the measured flux $\sim 1/km^2yr$ for cosmic rays of energy
$\ge 10^{19}~eV$.

\vspace{15pt}
\begin{center}
{\em III. Lifetime of Vortons}
\end{center}

We may estimate the decay probability of the vorton by a tunneling
expression through a barrier of height $\Delta E$ and width $\Delta R$

\begin{equation}
\tau_v ^{-1} \sim M_v~\exp(-\Delta E~\Delta R)~~~.
\end{equation}

It can be evaluated [9] that the vorton mass is $M_v \sim N \eta$ where $N
\sim Z$ is the topological or charge number. Its length is $L \sim N
\eta^{-1}$ and the maximum number of produced particles is around $N$.

For the case in which one considers that the loop collapses and disappears,
the barrier height $\Delta E$ corresponds to the energy which must be
supplied to cut the configuration across the area limited by the loop in
order to recover the same topology as the vacuum. One may expect therefore
$\Delta E \sim N^2 \eta$.  On the other hand the barrier width $\Delta R$
may be estimated by the contraction of the loop to a point, giving rise to
the $N$ free particles equivalent to the initial energy of the vorton. In
this way $\Delta R \sim N \eta^{-1}$.  If this is the dominant decay channel
the vorton lifetime

\begin{equation}
\tau_v ^{-1} \sim N~M_X~\exp(-N^3) \sim N~{10^{47} \over yr} \exp{(-N^3)}
\end{equation}

\noindent
would be so large that for any reasonable $N\ge 10$ Eq.(2) would give a
negligible cosmic ray flux.

But another decay mode is the one in which the vorton emits a
carrier conserving angular momentum [10]. To make a more detailed evaluation
of the lifetime along this line, the energy of a vorton of radius $R$ is

\begin{equation}
E=2 \pi R \mu +K \frac {N^2}{R}~~~,
\end{equation}

\noindent
where the first term comes from the normal string tension $\mu \sim \eta^2$
and the second one from the current and charge contributions $J^2+Q^2$ with
$K<1$ depending on the nature of the carriers. The minimization with respect
to $R$ gives

\begin{equation}
R^*={\sqrt {\frac{K}{2\pi \mu}}}~N~~~,~~~E^*=2\sqrt {2 \pi K \mu}~N~~~.
\end{equation}

Thinking on the simple case where along the string a charged field $\sigma$
is oscillating with amplitude $\sigma_0$ and a phase which changes in $2 \pi
N$ around the loop, the decay probability will correspond to Eq.(7) to pass
to a $N-1$ chiral string and one emitted particle. To evaluate the barrier
height one has to put $\sigma_0 \rightarrow 0$ along one wavelength
and extract one particle with momentum conservation, requiring therefore

\begin{equation}
\Delta E=\Delta V~\delta^2~\frac 1 \eta~+~\sqrt{m_\sigma^2+
({\frac N {R^*}})^2}~-~\frac N {R^*}~~~.
\end{equation}

The increase of potential in the core may be estimated to be
$\Delta V \approx m_\sigma^2 \sigma^2_0$ since its minimum is expected
to occur there for $|\sigma|=\sigma_0$, the string width
$\delta~>~\eta^{-1}$ for the superconducting case [11] and the momentum for
the massless carrier inside the string is $N/R^*$ because the uncertainty
principle must be considered for a segment $\sim \eta^{-1}$ both for
fermions and bosons condensate. The particle $\sigma$ acquires a mass
$m_\sigma \approx f \eta$ outside the core through coupling with the neutral
field $\phi$ responsible for the U(1) breaking which generates the string,
and keeps the same momentum as inside the core to conserve angular momentum.

The barrier width comes from the separation of the emitted
particle up to the position where, always conserving angular
momentum, the total energy of the configuration equals that of
the original string

\begin{equation}
\sqrt{m_\sigma^2+({\frac N {R^*+\Delta R}})^2}~+~2 \sqrt{2 \pi K \mu}~
(N-1)~=~2 \sqrt {2 \pi K \mu}~N~~~,
\end{equation}

\noindent
which allows to extract $\Delta R \propto N~\eta^{-1}$.

For the use of Eq.(7) we will have now $\Delta E~\Delta R \approx b~N$.
If $m_\sigma \sim 10^9 GeV$ it turns out [9] that $N \sim 10^6$ predicting
extremely stable loop. If instead $m_\sigma~\sim~M_{GUT}$ the number $N
\sim 10$ and with $b \sim 15$, which is perfectly possible because of the
contribution to $\Delta E$ of the first term of Eq.(11) with the expected
$\sigma_0~\sim~\eta$, the required value $\tau_v \sim a~few~10^{10}~yr$  can
be obtained to have from Eq.(6) with $N_c=10$ a flux of the order of one
event per $km^2$ of detector per year.

An objection regarding the use of Eq.(6) for vortons with coinciding scales
of string formation and superconducting condensate is that their density
should overcome the critical one. However, it must be considered that this
statement corresponds to neglecting their quantum decay. In addition, other
causes of decrease of vorton density may be the disappearance of zero modes
in phase transitions subsequent to their formation [12] and electromagnetic
selfinteractions [13]. Therefore, it is not unconceivable that these
small vortons are compatible with the universe critical density.

\vspace{12pt}
\begin{center}
{\em IV. Conclusions}
\end{center}

We have seen that small loops of superconducting strings which contain
around ten heavy carriers may decay by tunneling producing a flux of high
energy cosmic rays compatible with the expected one to be tested in
the future by observatories like the Auger Project. This corresponds to
coincident scales for formation of strings and superconducting condensate.
On the contrary larger loops which might have become superconducting at a
lower scale would be so stable that can be excluded as sources of UHECR.

To obtain a more precise prediction of the contribution of decaying vortons
two aspects should be refined. One of them is the study of detailed models
related to Grand Unified Theories which produce superconducting strings and
the calculation of their density taking into account quantum effects and the
enhancement or depletion due to the thermal history of the universe after
their formation. The other is the precise analysis of the vortons lifetime
going beyond the present semiclassical estimation, considering all the
involved fields including the GUT gauge bosons and identifying the
instantons responsible for the decay of these configurations.

\begin{center}
{\em  Acknowledgements}
\end{center}

We thank Diego Harari for useful discussions.

\begin{center}
{\bf REFERENCES}
\end{center}
\begin{enumerate}
\item Auger Design Report 1997.

\item Th.K. Gaisser {\it Cosmic Rays and Particle Physics}, Cambridge
University Press (Cambridge, 1990).

\item F. Halzen, astro-ph 9703004.

\item C.T. Hill, D.N. Schramm and T.P. Walker,
{\it Phys. Rev.} {\bf D36} (1987) 1007;\\
  P. Bhattacharjee and N.C. Rana,  {\it Phys. Lett.} {\bf B246} (1990) 365;\\
  G. Sigl,  {\it Space Sc. Rev.} {\bf 75} (1996) 375;\\
  G. Sigl, S. Lee, D.N. Schramm and P. Bhattacharjee,
{\it Science} {\bf 270} (1995) 1977;\\
  R.J. Protheroe and P. Johnson,  {\it Astropart. Phys.} {\bf 4} (1996) 253.

\item A.J. Gill and T.W.B.  Kibble {\it Phys. Rev.} {\bf D 50} (1994) 3660.

\item E. Witten  {\it  Nucl. Phys.} {\bf B249} (1985) 557.

\item R.L. Davis and E.P.S. Shellard, {\it Phys. Lett.} {\bf B 209} (1988)
485, and {\it Nucl. Phys.} {\bf B323} (1989) 209;\\
  B. Carter,  {\it Ann. N. Y.  Acad. Sci.} {\bf 647} (1991) 758;\\
  B. Carter and X. Martin,  {\it  Ann. Phys.} {\bf 227} (1993) 151;\\
  X. Martin and P. Peter, {\it Phys. Rev. } {\bf D51} (1995) 4092.

\item A.-C. Davis and P. Peter, {\it Phys. Lett.} {\bf B 358} (1995) 197.

\item R. Brandenberger, B. Carter, A.-C. Davis and M. Trodden,
{\it Phys. Rev.} {\bf D54} (1996) 6059.

\item R. L. Davis, {\it Phys. Rev.}{\bf D 38} (1988) 3722.

\item J. Ambjorn, N. Nielsen and P. Olesen, {\it Nucl. Phys.} {\bf B310}
(1988) 625.

\item S.C. Davis, A.-C. Davis and W.B. Perkins, hep-ph 9705464.

\item A. Gangui, P. Peter and C. Boehm, hep-ph 9705204.

\end{enumerate}
\end{document}